\documentclass[twocolumn]{aastex631}

\usepackage{chngpage}
\usepackage{booktabs}
\usepackage{amsmath}
\usepackage{bm}

\newcommand{\uvec}[1]{\boldsymbol{\hat{\textbf{\textit{#1}}}}}

\newcommand\RGMX{\bgroup\markoverwith{\textcolor{cyan}{\rule[0.5ex]{4pt}{1pt}}}\ULon}

\newcommand\ACCX{\bgroup\markoverwith{\textcolor{red}{\rule[0.5ex]{4pt}{1pt}}}\ULon}
\shorttitle{CBDs and interstellar asteroids}
\shortauthors{Childs \& Martin}
\graphicspath{{./}{figures/}}

\begin{document}

\title{Misaligned circumbinary disks as efficient progenitors of interstellar asteroids}

\author[0000-0002-9343-8612]{Anna C. Childs}
\author[0000-0003-2401-7168]{Rebecca G. Martin}
\affiliation{Nevada Center for Astrophysics, University of Nevada, Las Vegas, NV 89154, USA}
\affiliation{Department of Physics and Astronomy, University of Nevada, Las Vegas, 4505 South Maryland Parkway,
Las Vegas, NV 89154, USA}



\begin{abstract}
Gaseous circumbinary  disks (CBDs) that are highly inclined to the binary orbit are commonly observed in nature.  These disks harbor particles that can reach large mutual inclinations as a result of nodal precession once the gas disk has dissipated.    With $n$-body simulations that include fragmentation we demonstrate that misaligned disks of particles can be efficient progenitors of interstellar asteroids (ISAs). Collisions that take place between particles with large mutual inclinations have large impact velocities which can result in mass ejection, with a wide range of fragment sizes and ejection velocities.  We explore the binary parameters for which the majority of the terrestrial planet forming material is ejected rather than accreted into planets.  The misalignment required to eject significant material decreases with binary eccentricity. 
 If the distribution of binary eccentricity is uniform and the initial particle CBD orientation relative to the binary orbit is isotropic, about 59\% of binaries are more likely to eject the majority of their CBD terrestrial planet disk mass through high velocity body-body collisions rather than retain this material and build terrestrial planets.  However, binary--disk interactions during the gas disk phase with non-zero disk viscosity will reduce this fraction. The composition, small size, highly elongated shape, and tumbling motion of `Oumuamua is consistent with ISAs generated by  misaligned CBDs.
\end{abstract}

\keywords{Binary stars (154), Asteroids (72), Extrasolar rocky planets (511), Interstellar objects (52), Exoplanet formation (492)}

\section{Introduction} \label{sec:intro}

Circumbinary gas disks (CBDs) with large misalignments  relative to the binary orbital plane  are commonly observed in nature \citep[e.g.][]{Chiang2004, Kohler2011, Andrews2014,  Brinch2016, Fang2019, Takakuwa2017, Kennedy2019,Zhu2022,Kenworthy2022}. The degree of CBD misalignment often increases with binary separation and eccentricity \citep{Czekala2019}. Misalignments may initially arise as a result of turbulence in the molecular gas cloud \citep{Offner2010, Tokuda2014, Bate2012}, later accretion of material by the young binary \citep{Bates2010, Bate2018}, warping by a tertiary companion such as a stellar flyby \citep{Nealon2020} or, if the binary forms from a cloud whose elongated axis is misaligned to its rotation axis \citep{Bonnell1992}. 
The misaligned disk may precess as a solid body if the communication timescale is shorter than the precession timescale  \citep{PT1995,Larwood1996}.  As a result of dissipation, a viscous disk evolves towards either a coplanar or polar ($90^{\circ}$) alignment to the binary orbital plane \citep{Martin2017, Lubow2018, Zanazzi2018,Cuello2019} although depending on the binary and disk parameters, the timescale for alignment may be longer than the disk lifetime meaning that planet formation can take place in misaligned disks \citep[e.g.][]{Martin2018}.

The late stage of terrestrial planet formation takes place after Moon-size planetesimals and Mars-size embryos have formed and the gas disk has dispersed.  These solid bodies interact with one another through purely gravitational interactions to form terrestrial planets through core accretion \citep{Artymowicz1987,Lissauer1993, Pollack1996}.  Coplanar and polar circumbinary orbits are stationary states in which the particles do not undergo significant nodal precession.  As a result, terrestrial planets can efficiently form in coplanar \citep{Quintana2006,Childs2021} and polar aligned \citep{Childs2021ApJ} circumbinary disks through core accretion.  Such terrestrial circumbinary planets (CBPs) have yet to be observed however.  While this may be attributed to observational bias against such small planets in a circumbinary orbit \citep{Windemuth2019,MartinDV2021,Standing2022}, it may also indicate that terrestrial planets do not form through core accretion in a circumbinary disk or, that current core accretion models are missing key physics.

In a disk that is misaligned from a stationary state, nodal precession can lead to large mutual misalignments and collisions with large impact velocities that may result in ejection from the system.
Whether planets can form or not depends upon the misalignment and the binary eccentricity.
Terrestrial CBPs that do form end up either coplanar or polar to the binary orbital plane since mergers between bodies with random nodal phase angles leads to lower inclination to the stationary states \citep{ChildsMartin2022}.  While collisions were resolved with only perfect merging in \cite{ChildsMartin2022}, in this work we consider fragmentation as a more realistic outcome from such high energy collisions.  

`Oumuamua was the first confirmed interstellar asteroid (ISA) to be observed \citep{ChambersK2016}.  `Oumuamua does not exhibit comet-like features indicating its composition is more consistent with a refractory planetoid \citep{Jewitt2017, Ye2017, Meech2017}.  This ISA has an unexpectedly low velocity relative to the local standard of rest (LSR), $\sim10 \, \rm km \, s^{-1}$ \citep{Mamajek2017} and has a highly elongated shape \citep{Meech2017, Bolin2018}.  The elongated shape and tumbling motion of this body suggests that it was involved in a violent collision in its past and was sent tumbling in its parent planetary system, indicating that collisions of solid bodies in other planetary systems is not uncommon \citep{Drahus2017, Fraser2018}.  The currently observed mass of `Oumuamua is estimated to be about $10^{-17} M_{\oplus}$ however, if `Oumuamua is composed of entirely N$_{2}$ ice, it could have lost up to 92\% of its initial mass upon entering the solar system \citep{Desch2021, Jackson_2021}. \cite{Seligman_2020} proposed that if `Oumuamua contained a significant amount of H$_{2}$ ice, it was likely pancake shaped when it was near periapsis and \cite{Mashchenko_2019} found that the light curve is consistent with such a shape.  In these cases, the observed properties of `Oumuamua are not representative of its origins.  

Various formation scenarios for `Oumuamua and other ISAs have been proposed such as ejections from a system as a result of tidal interactions with a white dwarf \citep{Rafikov2018}, ejections of fragments from tidally disrupted planets by a dense member of a binary system \citep{Cuk2018}, and ejections of a comet-like planetesimal from giant planet interactions \citep{Raymond2018}. 
Binary stars have been suggested to dominate rocky body ejections from planetary systems over that from single stars \citep{Jackson2018}. Planetesimals are ejected when they migrate inside the stability limit of the binary although this may require the presence of other planets \citep{Fitzmaurice2022}. 

In this letter, we propose that \textit{shortly after a highly misaligned circumbinary gas disk dissipates, solid bodies undergo violent collisions and become a source for ISAs such as `Oumuamua}.  The highly inclined particles have no requirement to migrate close to the binary as ejections occur over a wide radial range.  In Section~\ref{sec:dynamics} we first conduct three-body simulations to show how the initial disk misalignment and the binary eccentricity affect the particle mutual inclinations, and thus impact velocities.  In Section~\ref{sec:N-body} we then conduct $n$-body studies of terrestrial CBP formation in highly misaligned CBDs and resolve collisions with fragmentation.   We closely follow the collisions and the fate of the ejected material to better understand the nature of ISAs that are generated from misaligned CBDs.  In Section~\ref{sec:ISA_formation} we discuss the implication of these results for ISAs.  Lastly, we conclude with a summary of our findings in Section \ref{sec:Conlusions}.

\section{Circumbinary particle dynamics}\label{sec:dynamics}

A particle in a circumbinary orbit around an eccentric binary can undergo two types of nodal precession depending upon its initial inclination. For low initial tilt, the orbit is circulating, meaning that the particle orbit precesses around the binary angular momentum vector. If the initial inclination is above the critical value, the orbit will be librating, meaning that it precesses about the binary eccentricity vector \citep{Verrier2009,Farago2010,Doolin2011,Aly2015}. 
The critical inclination depends upon the binary eccentricity and the angular momentum of the particle \citep{Martin2019,Chen2019}. In the test particle limit, the minimum critical inclination that separates circulating and librating orbits is given by,
\begin{equation}\label{eq:i_crit}
    i_{\rm crit}=\rm{sin^{-1}} \sqrt{\frac{1-\textit{e}_{\rm b}^2}{1+4\textit{e}_{\rm b}^2}}
\end{equation}
\citep{Farago2010}. This critical inclination occurs for longitude of ascending node of $\phi=90^\circ$ measured in the frame of the binary \citep[see equation~(3) in][]{Chen2019}.

We measure the particle misalignment with respect to one of the two stable configurations, coplanar or polar. The inclination of the particle orbit relative to the binary orbit is given by
\begin{equation}
    i_{\rm b} =\textrm{cos}^{-1} (\uvec{l}_{\rm b} \cdot \uvec{l}_{\rm p}),
\end{equation}
and the inclination of the particle orbit relative the binary eccentricity vector is given by
\begin{equation}
    i_{\rm e}=\textrm{cos}^{-1} (\uvec{e}_{\rm b} \cdot \uvec{l}_{\rm p}),
\end{equation}
where $\bm{l}_{\rm b}$ and $\bm{l}_{\rm p}$ are the angular momentum vectors of the binary and particle, respectively,  $\bm{e}_{\rm b}$ is the eccentricity vector of the binary, and $\uvec{}$ denotes a unit vector.  For orbits with $\phi=90^\circ$ initially, if the initial particle inclination is smaller than the critical inclination (circulating orbit), we measure $i_{\rm b}$ and if the particle inclination is larger than the critical inclination (librating orbits) we measure $i_{\rm e}$.

\begin{figure*}
\centering
	\includegraphics[width=2\columnwidth]{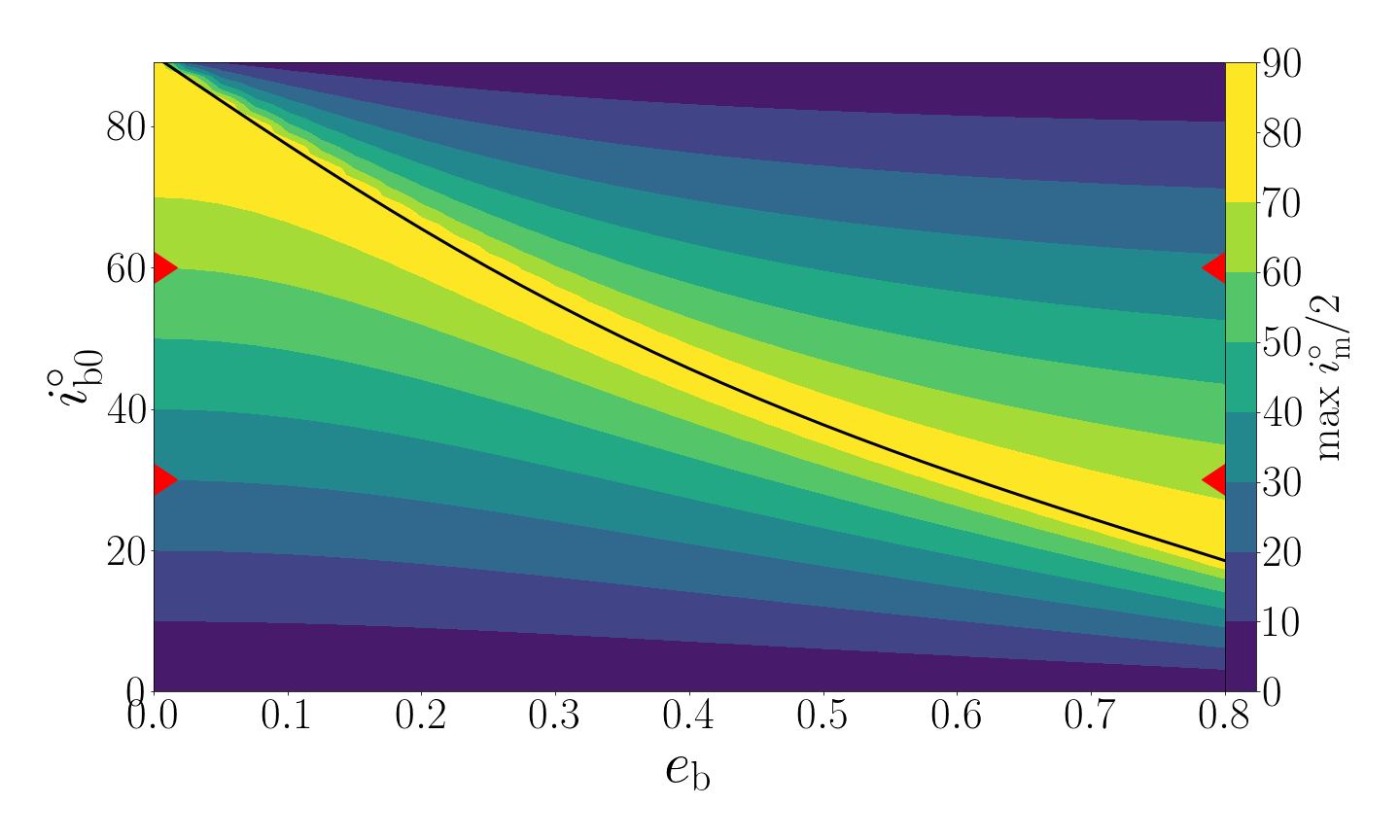}
    \caption{Contour plot showing the maximum inclination a particle reaches relative to the axis about which it precesses (half the maximum mutual inclination $i_{\rm m}/2$) around binaries with different eccentricities and initial inclination to the binary, $i_{\rm b 0}$ with initial longitude of ascending node $\phi=90^\circ$.  The black solid line shows the critical inclination for the various binary eccentricities (Equation~\ref{eq:i_crit}).  For circulating orbits (below the black line) we show max$\,i_{\rm b}$ while for librating orbits (above the black line) we show max$\,i_{\rm e}$.  The red triangles mark the binary parameters used in our $n$-body simulations in Section~\ref{sec:N-body}.}
    \label{fig:countour_eb}
\end{figure*}

The maximum impact velocity for a collision between two particles orbiting at the same semi-major axis in circular orbits with Keplerian velocity, $v_{\rm K}$, can be estimated with 
\begin{equation}\label{eq:vel}
   v_{\rm max}=2\textit{v}_K\rm sin \left ( \textit{i}_{\rm max}/2 \right ) ,
\end{equation}
where $i_{\rm max}={\rm max\,} i_{\rm m}$ is the maximum value over a nodal precession period of the mutual inclination  between the two particles, $i_{\rm m}$.  In a colliding system with particles of different nodal precession rates, $i_{\rm max}$  is twice the maximum inclination a particle reaches over its nodal precession period measured with respect to the stationary inclination about which it precesses (either coplanar or polar).

To probe the maximum mutual inclinations expected in a circumbinary disk as a function of binary eccentricity we conduct three-body simulations of a very low mass particle at an orbital radius of $5 \, a_{\rm b}$ from the barycenter of the three-body system.  We change the eccentricity and initial inclination of the particle to the binary, $i_{\rm b0}$, and integrate for two full nodal precession periods (see Equations 6-10 of \cite{ChildsMartin2022}) using the WHFAST integrator in \textsc{rebound} \citep{Rein2012}.  Initially the longitude of ascending node is $\phi=90^\circ$ in all cases. 

Figure \ref{fig:countour_eb} shows the results from our three-body simulations.  We plot  the maximum inclination a single particle experiences relative to the axis about which it precesses (coplanar or polar), which is equivalent to $i_{\rm m}/2$, as a function of the particle's initial inclination to the binary, $i_{\rm b0}$, and binary eccentricity, $e_{\rm b}$.  We plot the critical inclination (Equation~\ref{eq:i_crit}) in the solid black line.  We see that as the binary eccentricity increases, the less inclined the particle needs to be to reach the maximum inclination away from a stable configuration. This is expected since the maximum inclination a particle experiences is near the critical inclination, which decreases as binary eccentricity increases.  This indicates that CBDs with relatively modest inclinations around highly eccentric binaries can harbor particles with high mutual inclinations and thus, colliding particles can experience high impact velocities.  We do not expect this trend to change for binary systems with different mass ratios or different separations since these effects change only the timescale for the particle dynamics. The critical inclination and the particle dynamics depend only on the binary eccentricity and angular momentum ratio of the particle to the binary \citep{Farago2010,Martin2019,Chen2019}.

\begin{table*}
\caption{The model name, binary eccentricity ($e_{\rm b}$), initial inclination above the binary plane ($i_{\rm b0}$), initial inclination away from a polar configuration ($i_{\rm e0}$), and the particle surface density fit ($\Sigma$) \citep[from Figure 2 in][]{Childs2021ApJ} used for our $n$-body simulations.  We denote whether the particle orbits in the disk are initially circulating (C) or librating (L).  We list the multiplicity of the terrestrial planetary system and the average and standard deviation of the planet properties.  A planet is defined as a body with mass $M_{\rm p}\ge M_{\oplus}$.  In the last column, we list the mean and standard deviation for the fraction of disk mass that is ejected in each run.
} 
\begin{adjustwidth}[]{.5cm}{}
\resizebox{1\linewidth}{!}{
\hskip-4.0cm
\begin{tabular}{c|c|cccc|ccccc|c}
  \hline
{Model}  & {$e_{\rm b}$} & {$i_{\rm b0}^{\circ}$} & {$i_{\rm e0}^{\circ}$} & {$\Sigma$} & {C}/{L} & {\#} &  {$M_{\rm p}/M_\oplus$} &  {$a_{\rm p}/ \rm au$} &  {$e$} &  {$i_{b/ \rm e}^{\circ}$} &  {$M_{\rm e}/M_{\rm d}$} \\
\hline
C30 & 0.0  & 30.0  & 60.0 & CC & C & 1.82 $\pm$ 0.48
& 1.76 $\pm$ 0.55
&  2.25 $\pm$ 0.38
& 0.05 $\pm$ 0.03
& 4.97 $\pm$ 2.64 & 0.07 $\pm$ 0.05\\
C60 & 0.0  & 60.0  & 30.0 & CC & C & 0 & - &- &- &-& 0.81 $\pm$ 0.21\\
E30 & 0.8  & 30.0  & 60.0 & EP & L & 0 & - &- &- &-& 0.79 $\pm$ 0.22\\\
E60 & 0.8  & 60.0  & 30.0 & EP & L & 1.62 $\pm$ 0.57
& 1.93 $\pm$ 0.66
& 2.32$\pm$ 0.48
& 0.07 $\pm$0.04
& 4.51 $\pm$2.73& 0.13 $\pm$ 0.06\\

\hline
\end{tabular}
}
\end{adjustwidth}
\label{tab:systems}
\end{table*}

\section{Terrestrial circumbinary planet formation}\label{sec:N-body}
\begin{figure*}
\centering
	\includegraphics[width=1\columnwidth,height=0.5\textheight]{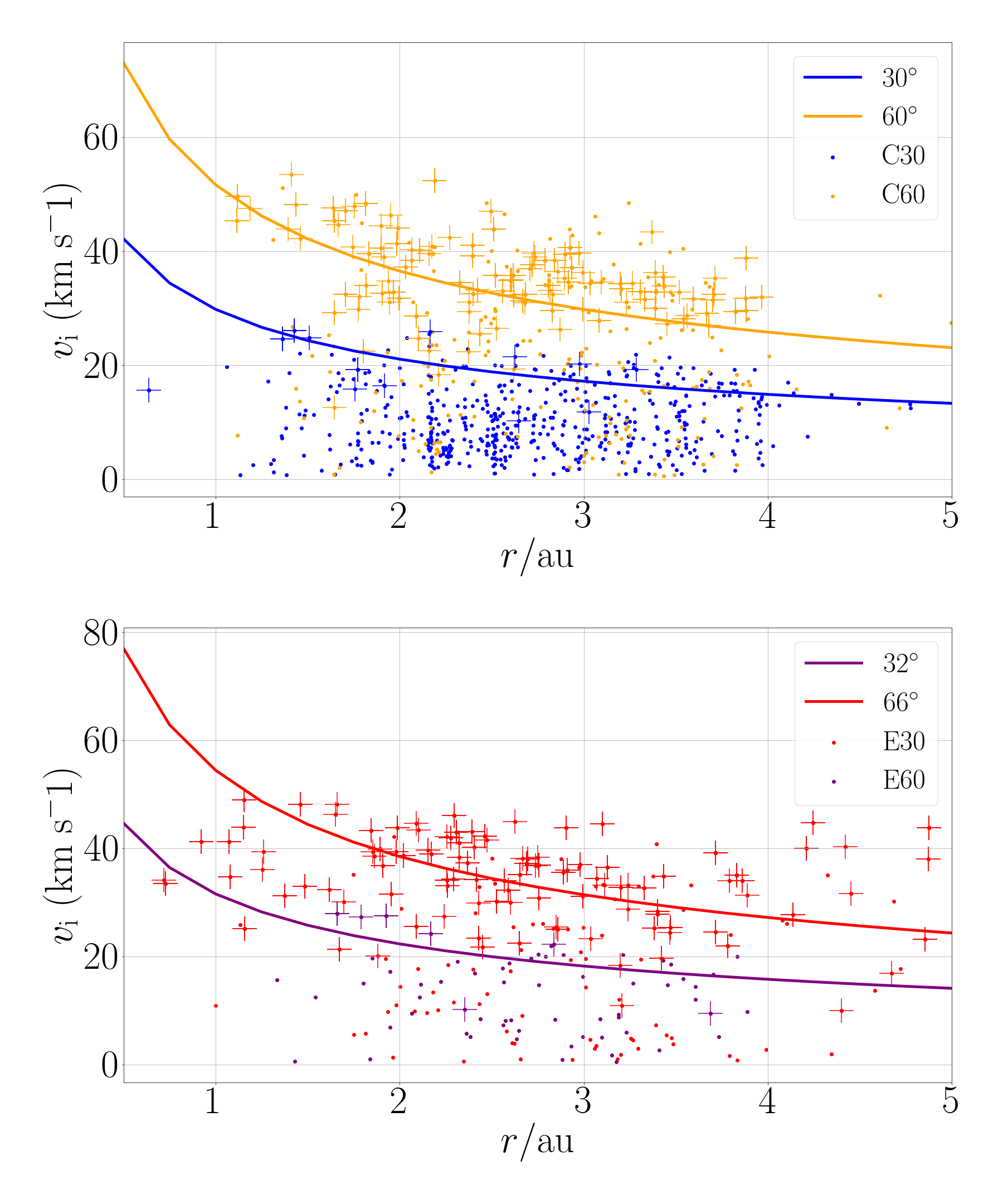}
	\includegraphics[width=1\columnwidth,height=0.5\textheight]{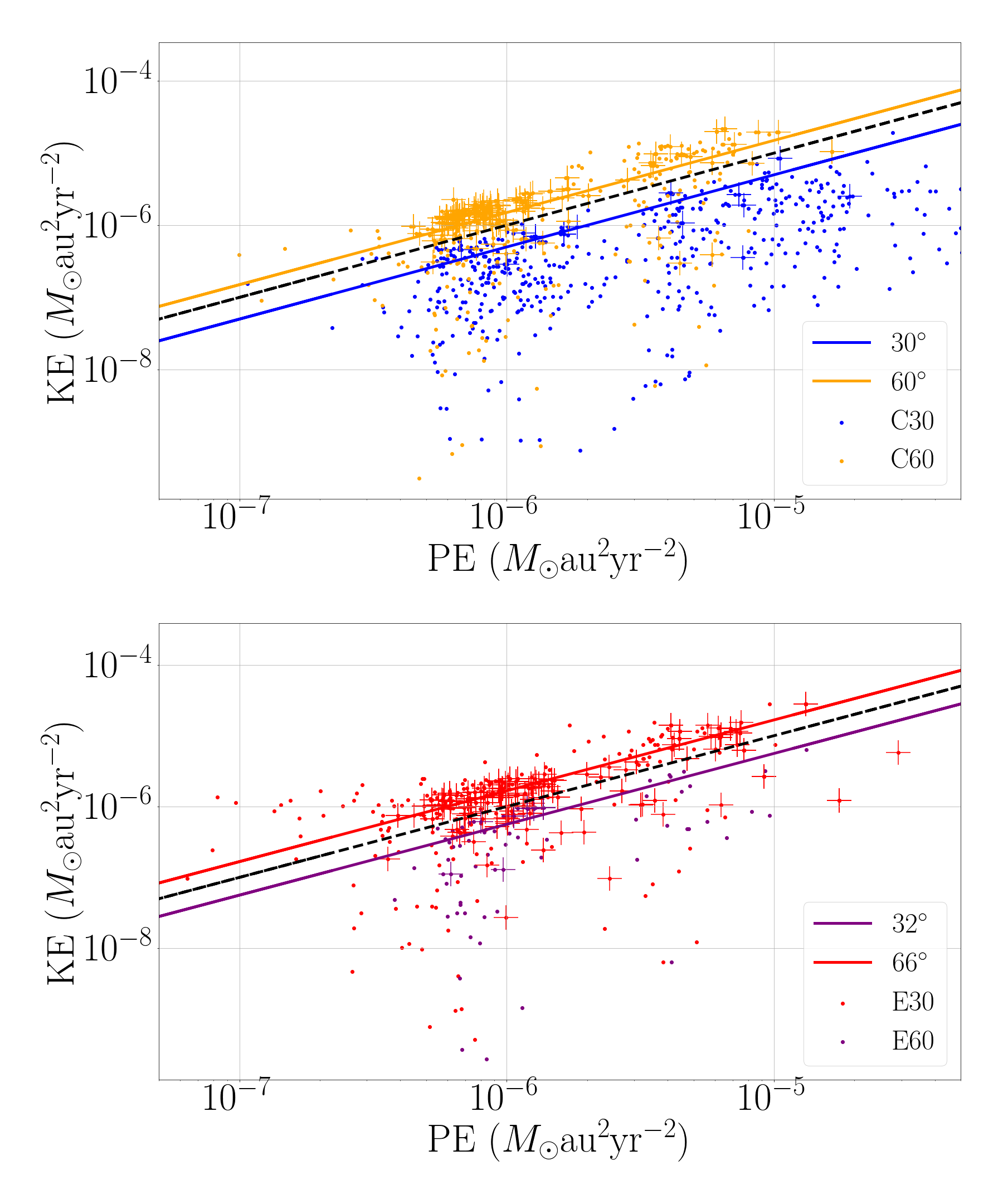}
    \caption{The left panels show the impact velocity versus the semi-major axis for all the collisions that took place in one run of each model.  The pluses mark collisions that lead to ejections.  We also plot the curves that represent the circular orbit maximum impact velocity, Equation \ref{eq:vel}, for each model.  The right panels show the KE and PE of each collision and the KE/PE lines using the circular orbit maximum impact velocity for each model.  The KE=PE line is marked by a black dashed line.}
    \label{fig:v_semi}
\end{figure*}

We now explore simulations of terrestrial planet formation in the inner parts of an initially misaligned circumbinary disk including the effects of fragmentation. The fragmentation code we use is detailed in \cite{ChildsSteffen2022}.  We distribute 26 Mars-sized embryos ($m \approx 0.1 \, M_{\oplus}$) and 260 Moon-sized planetesimals ($m \approx 0.01 \, M_{\oplus}$) along the fits from SPH simulations detailed in \cite{Childs2021}.   This bimodal mass distribution is adopted from $n$-body simulations that were successfully able to recover the masses of the solar system terrestrial planets \citep{Chambers2001}. 

Unlike WHFAST, the IAS15 integrator in \textsc{rebound} is a high-precision non-symplectic integrator that is able to resolve close encounters and collisions \citep{Rein2015}.  This feature is necessary for modeling core accretion of the terrestrial planets.  To overcome the excessive CPU time associated with a non-symplectic integrator we apply an expansion factor of $f=25$ to all the particles after integrating the system for $100 \, \rm Kyr$ and the phase angles of the particles have randomized.  \cite{Childs2021} performed convergence tests with $f=25$ and smaller expansion factors in $n$-body simulations of CBP terrestrial planet formation using the IAS15 integrator.  They found that while larger expansion factors lead to some differences in system architecture, the general planet formation trends that emerge as a function of binary eccentricity and separation remain.  Furthermore, \cite{ChildsSteffen2022} studied the effects of expansion factors with fragmentation.  Their findings indicate that our use of an expansion factor will lead to shorter collision timescales and more damped orbits which are more likely to underestimate impact velocities in collisions.

We set the minimum fragment mass to half the size of the Moon ($m \approx 0.005 \, M_{\oplus}$).  Fragments are expected to be much smaller, but we choose this value to reduce the CPU time of the simulations.  While this fragment mass is orders of magnitude larger than `Oumuamua, it should be viewed as an upper limit as fragment producing collisions will produce a wide distribution of fragment sizes \citep{Leinhardt2012}.  

The orbital elements for each body are randomly chosen in each run.  We consider perfectly circular binaries with $e=0.0$ and highly eccentric binaries with $e=0.8$ with particle disks that are initially inclined $30^{\circ}$ and $60^{\circ}$ above the binary orbital plane.  A particle is considered ejected from the system once its distance from the barycenter of the system exceeds $100 \, \rm au$.  The binary consists of equal mass stars with a total binary mass of $1 \, M_{\odot}$ separated by $0.5 \, \rm au$.  We perform 50 runs for each setup and integrate for a total of $7 \, \rm Myr$. 

The different binary models and their corresponding binary eccentricities are listed in Table \ref{tab:systems} and are marked by red triangles in Fig.~\ref{fig:countour_eb}.  The initial surface density profile is taken to be that of steady circumbinary gas disk as described in \cite{Childs2021ApJ}. At least initially, both particle disks around the circular orbit binary are in circulating orbits while around the eccentric orbit binary they are librating.

Our simulations reveal that the C60 and E30 eject the most material and do so throughout the entirety of the simulation.  The sustained ejection rates indicate that particles are not just being quickly ejected in the inner, unstable region of the disk close to the binary as a result of strong binary-particle interactions \citep[e.g.][]{Holman1999,Chen2020}.  Particles found at larger, initially more stable, orbits also get ejected on longer timescales as a result of particle-particle interactions.  On average, the C60 and E30 systems eject $~80\%$ of their disk mass and the C30 and E60 systems eject $~1\%$ of their disk mass by the end of our simulations.  The large difference between these ejection percentages is the result of the different mutual inclinations the particles reach in the runs.  In Figure \ref{fig:countour_eb} we see that the C60 and E30 particles can reach maximum mutual inclinations of $i_{\rm m} = 120^{\circ}$ and $i_{\rm m} = 132^{\circ}$, respectively, for circular orbits.  Such large mutual inclinations will result in high impact velocities when a collision takes place which is likely to result in mass ejection.  The C30 and E60 particles reach maximum inclinations of $i_{\rm m} = 60^{\circ}$ and $i_{\rm m} = 64^{\circ}$, respectively, which will result in much lower impact velocities and less mass being ejected from the system.  

Because of the high ejection rates in C60 and E30, no terrestrial planets are formed.  On average, the C30 and P60 systems form at least one terrestrial planet with a mass greater than $1 \, M_{\oplus}$ that is nearly circular and coplanar to the circular binary or polar to the eccentric binary.

\begin{figure*}
\centering
	\includegraphics[width=1.8\columnwidth]{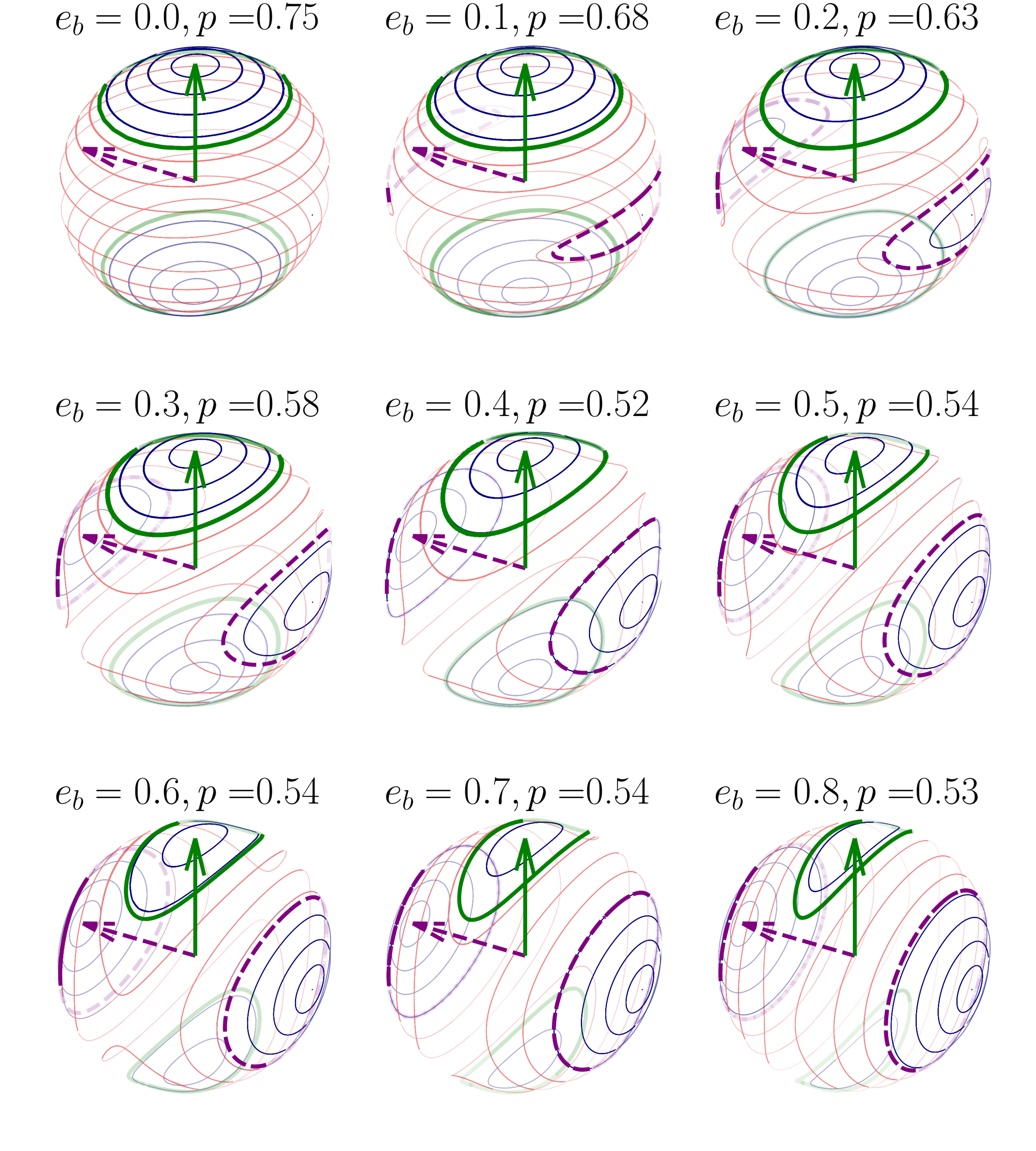}
    \caption{
    Precession paths for the angular momentum vector of the particle, $\bm{l}_{\rm p}$, in the frame of the binary for nine different binary eccentricities. The solid green and dashed purple arrows represent the angular momentum and eccentricity vectors of the binary, respectively.  The dark blue curves represent orbits with inclinations that are always less than $45^{\circ}$ from either coplanar or polar and the light red curves represent orbits with inclinations that may be more than $45^{\circ}$.  The bold green and dashed purple curves mark the boundaries between each region, where the particle inclination just reaches $45^{\circ}$.}
    \label{fig:phase_plots}
\end{figure*}

The final planetary systems with fragmentation are similar to the final planetary systems of \cite{ChildsMartin2022} who modeled planet formation in similar misaligned CBDs but resolved collisions with only perfect merging.  They did not consider a system analogous to E60, but our E60 planetary systems closely resemble those formed in the C30 runs since both are inclined by $30^\circ$ to a stationary inclination.  The planet eccentricities and inclinations are slightly damped, relative the planets formed with only perfect merging, due to dynamical damping from the fragments.  A notable difference between the planetary systems that formed with perfect merging and fragmentation is the formation timescale.  In agreement with \cite{Quintana2016} who compared planet formation in systems with and without fragmentation, fragmentation approximately doubles the formation and CPU time but results in similar planetary systems as those formed with only perfect merging.

\section{Formation of interstellar asteroids}\label{sec:ISA_formation}

The left column of Figure \ref{fig:v_semi} shows the impact velocity versus the semi-major axis for all the collisions that took place in one run of each model.  The pluses mark collisions that lead to ejections.  We also plot the curves that represent the circular orbit maximum impact velocity, Equation \ref{eq:vel}, for each model.  We use the maximum inclination the particle reached in our three-body simulations for $i_{\rm m}/2$, to predict the maximum impact velocity in systems with the same binary setup.  We see that the majority of the collisions in each model are near the circular orbit impact velocity curves.  Particles that are on eccentric orbits can  result in larger impact velocities than $v_{\rm max}$.  We see that collisions which lead to ejections can be found for a wide range of velocities but are typically found when the impact velocity is greater than $\sim 15 \, \rm km \, s^{-1}$.

When the kinetic energy (KE) of a body is equal to or less than the potential energy (PE) of the body, it will remain gravitationally bound to the binary.  The right column of Figure \ref{fig:v_semi} shows the KE=PE line for a range of masses and radii with a black dashed line.  We expect collisions on and below this line to remain bound to the binary and collisions above this line to be ejected from the system.  This line also corresponds to $v_{\rm i}=2v_{\rm k} \rm sin 45^{\circ}$.  Using data from one run in each setup, we plot the PE and KE of each collision using the total mass of the colliding system, the impact velocity, and the last recorded semi-major axis of the target.  Collisions that involve bodies that are eventually ejected from the system are marked with a plus.  We also plot the line with slope $\frac{KE}{PE}$ using the circular orbit maximum impact velocity for each system, over a range of radii and mass.  CBDs with particles $45^{\circ} \leq i_{\rm m}/2$ will result in collisions where $1 \leq \frac{KE}{PE}$, which will lead to ejection from the system unless dynamics from the multi-body system prevent such.  We see that the KE versus PE line for the C60 and E30 system, which have a circular orbit maximum impact velocity of $v_{\rm i}=2 v_{\rm k} \rm sin 60^{\circ}$ and $v_{\rm i}=2 v_{\rm k} \rm sin 66^{\circ}$ respectively, is above the KE=PE line where $\sim 80 \%$ of ejections are found.  The C30 and E60 systems, which have a circular orbit maximum impact velocity of $v_{\rm i}=2 v_{\rm k} \rm sin 30^{\circ}$ and $v_{\rm i}=2 v_{\rm k} \rm sin 32^{\circ}$ respectively, is found below the KE=PE line explaining the low number of ejections we observe in these systems.  These $i_{\rm m}/2$ values are taken from the corresponding binary systems in Figure \ref{fig:countour_eb} which are marked by red triangles.

The largest remnant of a collision that results in fragmentation is $M_{\rm lr}$.  The formula for calculating $M_{\rm lr}$ is taken from \cite{Leinhardt2012} and is a function of impact energy and the impact angle.  As expected, the systems that are inclined $60^{\circ}$ away from a stable configuration (C60 and E30) experience collisions with the highest impact velocities which result in smaller values for the $M_{\rm lr}$.  Although the resolution of our simulations is limited by the minimum mass of the fragment we define, we calculate the true $M_{\rm lr}$ for all collisions.  The smallest mass of the largest remnant calculated in our simulations (although not included in the simulation due to lower limits on the fragment mass) is $2 \times 10^{-6} \, M_{\oplus}$.  While this is still orders of magnitudes larger than the expected mass of `Oumuamua, this is the largest remnant expected from a collision which will also produce a distribution of smaller fragments.  The most massive body ejected was $0.91 \, M_{\oplus}$ in an E60 run and so, we expect a large distribution of fragment masses in misaligned CBDs.

Using our results, we can place an upper limit on the fraction of binaries in the galaxy that are likely to eject the majority of their terrestrial planet building material.  
\cite{Moe2017} compiled observations of early-type binaries to quantify the distributions of binary eccentricities.  Close-in binaries with separations $\leqslant 0.4 \, \rm au$ have small eccentricities $\leqslant0.4$ due to tidal circularization, while more widely separated binary eccentricities are weighted to larger values.  However, here we assume that binary eccentricity is uniformly distributed in the range 0.0 - 0.8 \citep{Raghavan2010} and that the initial orientation of the CBD is uniformly distributed for simplicity.

A CBD ejects most of its solid material when the mutual inclination between colliding bodies becomes greater than $90^{\circ}$ which is $v_{\rm i}=2v_{\rm k} \rm sin 45^{\circ}$, the PE=KE black line in Figure \ref{fig:v_semi}.  The probability, $p$, that a binary ejects most of its solid material can be estimated with the fraction of orbits that a particle in a CBD around the binary has a maximum inclination to either coplanar or polar that is greater than $45^{\circ}$.

Figure \ref{fig:phase_plots} shows precession paths of a particle angular momentum vector in the frame of the binary from our three-body simulations, for nine different binary eccentricities.  
To find $p$ we find the fraction of the sphere's surface area which corresponds to paths with particle inclinations that are at some point greater than $45^{\circ}$, given by
\begin{equation}
    p = 1 - \frac{(2A_{\rm e1} + 2A_{\rm e2})}{4\pi},
\end{equation}
 where  $A_{\rm e1}$ is the surface area of a bold green curved ellipse 
 that corresponds to the phase space that the particles inclination is always less than $45^{\circ}$ to coplanar and $A_{\rm e2}$ is the surface area enclosed by the bold and dashed purple curved ellipse which corresponds to the phase space where the particles inclination is always less than $45^{\circ}$ to polar.  To calculate the curved area of the ellipses we consider the edge as the cross section of an elliptical cylinder with the sphere. The surface area of the curved ellipse is then given by
\begin{equation}
    A_{\mathrm e} = 2 \pi a^2 - 4 a^2 \textrm{sin}^{-1} \left ( \frac{\sqrt{ a^2 + b^2}}{a} \right ),
\end{equation}
where $a$ and $b$ are the semi-major and minor axes, respectively, of the elliptical cylinder. 

In Figure \ref{fig:phase_plots} we list the $p$ value for each binary eccentricity.  Since we assume a uniform distribution of binary eccentricity, we take the average of all the $p$ values to find the fraction of binaries in the galaxy that are likely to eject most of their disk material.  We find $\Bar{p}=0.59$ meaning that with these assumptions, more than half of the binaries in the galaxy are more likely to eject their terrestrial planet disk mass, and produce ISAs, rather than form terrestrial planets.  Additionally, we find that binaries with eccentricities greater than 0.4 have the same probability, $p=0.54$, of ejecting most of their disk mass.

This is a strict upper limit to the fraction because we have assumed an initially isotropic orientation of the particle disk which is equal to the gas disk orientation at the time of disk dispersal. However, the gas disk may evolve towards either coplanar or polar alignment for non-zero disk viscosity. Depending on the binary and disk parameters, the alignment timescale can be longer than the disk lifetime and so the planetesimal disk may form in a  misaligned disk. We note that the two currently observed polar circumbinary gas disks have external companions \citep{Kennedy2019,Kenworthy2022} that truncate the outer part of the circumbinary disk leading to a radially narrow disk and a short alignment timescale \citep{Martin2022}. The timescale for gas disk alignment also depends upon a number of other binary and disk parameters including the binary semi-major axis, binary eccentricity, disk viscosity and disk temperature \citep[e.g.][]{Lubow2018}. Gas disk alignment doesn't proceed in a strictly linear fashion and even a small initial misalignment can lead to very large misalignments during the disk evolution \citep{Smallwood2019}. Because of the complexity of this problem, we leave a more detailed investigation of this to future work.

\section{Conclusions}\label{sec:Conlusions}
We conducted a suite of $n$-body simulations around circular and highly eccentric, equal mass binaries separated by $0.5 \, \rm au$.  The circumbinary particle disks were inclined by either $30^{\circ}$ or $60^{\circ}$ above the binary orbital plane.  We resolved all collisions with fragmentation which allowed us to analyze the collisions and better understand the post-collision mass and velocity distributions.  

We found that around highly eccentric binaries, CBDs with mild initial misalignment can result in large mutual inclinations between the particles.  The impact velocity between two bodies in circular Keplerian orbits with mutual inclination $i_{\rm m}$ is given by $v_{\rm i}=2 v_{\rm k} \mathrm{sin} (i_{\rm m}/2)$.  CBDs that harbor particles with mutual inclinations greater than $90^{\circ}$ have particle collisions with kinetic energies greater than the potential energy between the colliding system and the central binary which results in ejection from the system, unless dynamics from the multi-particle disk inhibit such.   This mechanism is an efficient source for ISAs with a wide range of sizes and velocities.  These ISAs will have characteristics consistent with a violent past, such as the small size, elongated shape, and tumbling motion of `Oumuamua.  ISAs formed in this way will mostly be rocky in composition since the terrestrial planets are expected to form inside of the snow line radius.

Assuming a uniform distribution of binary eccentricity and an isotropic distribution of the disk orientation relative to the binary, we find that 59\% of binaries in the galaxy are more likely to eject their CBD terrestrial material through high velocity particle-particle collisions rather than retain their material and build terrestrial planets. This is an upper limit to the fraction since a non-zero disk viscosity during the gas disk phases causes alignment towards either coplanar or polar alignment depending upon the initial misalignment.  These findings can help place constraints on occurrence rates for both ISAs and terrestrial CBPs.

\begin{acknowledgements}
We thank the anonymous referee for useful comments that improved
the manuscript.  We thank Stephen Lepp, Charlie Abod and Ian Rabago for help with Figure 3. We acknowledge support from NASA through grant 80NSSC21K0395.  Computer support was provided by UNLV’s National Supercomputing Center.

\end{acknowledgements}

\bibliography{ref}{}
\bibliographystyle{aasjournal}



\end{document}